\documentclass[aps,prb,twocolumn,groupedaddress,longbibliography,showpacs]
{revtex4-1}
\usepackage{graphicx}
\usepackage{amsmath,amsfonts,amssymb}
\usepackage{eucal}
\usepackage{bm}
\usepackage{url}
\begin{document}
\title{Projection operator approach to unfolding supercell band structures}
\author{M. Farjam}
\email[]{mfarjam@ipm.ir}
\affiliation{School of Nano-Science,
Institute for Research in Fundamental Sciences (IPM),
P.O. Box 19395-5531, Tehran, Iran}
\date{\today}
\begin{abstract}
  While the methodology of band structure unfolding
  has appeared in several publications,
  the original derivations of the unfolding formulas
  can be considerably simplified by using the $k$-projection method.
  In this work,
  more transparent derivations of unfolded spectral weights
  are given by using the projection operator approach.
  A range of illustrative examples are also presented
  which include finite and random one-dimensional chains,
  and Kekul\'e-textured honeycomb lattice.
\end{abstract}
\maketitle
\section{Introduction}
The band structure property of energy spectra
in a crystal depends on the possiblity of describing the system
in terms of a primitive unit cell.
In many electronic structure calculations, however,
it is necessary to use a supercell (SC) description,
which results in folded,
more densely packed and hard to interpret,
bands in the smaller supercell Brillouin zone (SBZ).
A challenging question is, therefore, how to recover,
from the supercell calculation alone,
the band structure in the larger primitive Brillouin zone (PBZ).
An unfolding procedure must reproduce in the larger PBZ
the original bands for a perfect system,
or give an effective band structure (EBS)
for an imperfect one.
The EBS has great interpretative value as it is comparable to
the band structure of the perfect system and can reveal the
perturbations caused by disorder.
The EBS is also important because it corresponds to
angle-resolved photoemission spectroscopy (ARPES) measurements.
\cite{grioni2005}

The methodology of band structure unfolding has been developed
in a number of publications for different representations of
the electronic structure.
\cite{boykin2005, boykin2007, ku2010, popescu2012}
The basic idea is to uncover the inherited symmetry of an
SC eigenfunction by comparing it with a reference Bloch function
of the underlying lattice.
A key idea that facilitates the conceptual description is to
express the band structure in terms of the spectral function.
\cite{ku2010, popescu2012}
Suitable unfolded spectral-weight expressions that serve as unfolding formulas
have been derived for plane-waves,
\cite{popescu2012}
the tight-binding method or Wannier functions,
\cite{boykin2007, ku2010}
and linear combination of atomic orbitals (LCAO).
\cite{lee2013}
A few studies have circumvented tedious derivations of these formulas by
defining the spectral weight formulas based on heuristic reasoning.
\cite{haverkort2011, deretzis2014}

Various recent papers have applied group theory
to band structure unfolding.
\cite{allen2013, *allen2013erratum, huang2014, tomic2014, chen2014}
This approach, which has been referred to as the $k$-projection method,
\cite{chen2014}
has the virtue that it does not require
an artificial reference system,
as it generates its own.
Nevertheless, it leads to the same results as the previous approach
based on the analysis of the spectral function.
\cite{allen2013}

Many other studies are using, and generalizing, the unfolding method to
interpret the electron and phonon spectra
of alloys, surfaces and disordered solids.
\cite{farjam2014, medeiros2014, rubel2014, bianco2014, medeiros2015,
berlijn2014, cahangirov2014, brommer2014, boykin2014}

The purpose of this paper is to present simplified, more transparent,
derivations of unfolding formulas based on
the $k$-projection operator.
Following the theoretical description in Sec.~\ref{theory},
a wide selection of toy models are also presented
in Sec.~\ref{results}
that illustrate various applications of band structure unfolding.

\section{\label{theory}Theory}

\subsection{The projection operator}
A convenient notation
is to use upper and lower case letters for SC and PC variables, respectively.
A supercell is considered to contain a number $\mathcal{N}$ of
primitive unit cells at positions $\mathbf{r}_j$.
Correspondingly, a set of $\mathcal{N}$
reciprocal lattice vectors of the SC system,
$\mathbf{G}_i$, unfold a wavevector
$\mathbf{K}$ in the SBZ onto wavevectors
$\mathbf{k}=\mathbf{K}+\mathbf{G}_i$ in the PBZ.
Using the translation operator, defined by
$\mathcal{T}(\mathbf{r})\psi(\mathbf{x})=\psi(\mathbf{x}+\mathbf{r})$,
it was shown in Ref.~[\onlinecite{allen2013}] that the operator
\begin{equation} \label{po}
  \mathcal{P}_{\mathbf{K} + \mathbf{G}_i} = \frac{1}{\mathcal{N}}
  \sum_{j=1}^{\mathcal{N}} \mathcal{T}(\mathbf{r}_j)
  e^{-i(\mathbf{K} + \mathbf{G}_i)\bm{\cdot}\mathbf{r}_j}
\end{equation}
projects out of a supercell eigenfunction with wavevector $\mathbf{K}$
a Bloch function with wavevector $\mathbf{k}=\mathbf{K}+\mathbf{G}_i$.
Equation~(\ref{po}) can also be obtained from the definition
of projection operator in group theory,
\begin{equation}
  \mathcal{P}_\mathbf{k} = \frac{1}{\mathcal{N}} \sum_{j=1}^\mathcal{N}
  \Gamma^{(\mathbf{k})}[\mathcal{T}(\mathbf{r}_j)]^\ast
  \,\mathcal{T}(\mathbf{r}_j),
\end{equation}
where $\Gamma$ is an irreducible representation of the group,
which is $e^{i\mathbf{k}\bm{\cdot}\mathbf{r}}$ for the translation group.
\cite{tinkham}
The projection operator is both Hermitian and idempotent.
An elementary proof that Eq.~(\ref{po}) has the
desired projective property is given in the following section.

\subsubsection*{Spectral weight}
The norms of the projected components of an eigenstate,
generally in the $[0,1]$ interval,
form the spectral weights of the unfolding procedure,
\begin{equation} \label{sw}
  W_{\mathbf{K}J}(\mathbf{G}_i) =
  \langle\mathcal{P}_\mathbf{k}\psi_{\mathbf{K}J}|
  \mathcal{P}_\mathbf{k}\psi_{\mathbf{K}J}\rangle =
  \langle\mathbf{K}J|\mathcal{P}_\mathbf{k}|\mathbf{K}J\rangle.
\end{equation}

\subsection{Plane-wave representation}
To prove that Eq.~(\ref{po}) is the required projection operator,
two conditions must be satisfied.
First, the operator must project out of
an SC wave function a PC Bloch function
and, second, it must be idempotent.
The proof below is based on a plane-wave representation.

An SC eigenfunction can be represented
as an expansion in plane waves,
\begin{equation} \label{pw}
  \psi_{\mathbf{K}}(\mathbf{x}) =
  \sum_\mathbf{G} C_{\mathbf{K}-\mathbf{G}}
  e^{i(\mathbf{K}-\mathbf{G})\bm{\cdot}\mathbf{x}},
\end{equation}
where the sum is over the set of all reciprocal lattice vectors
of the SC system.
\cite{ashcroft}
It is easy to verify the Fourier relation,
\begin{equation} \label{Fr}
  \frac{1}{\mathcal{N}} \sum_{j=1}^\mathcal{N}
  e^{-i(\mathbf{G}+\mathbf{G}_i)\bm{\cdot}\mathbf{r}_j}
  = \sum_\mathbf{g} \delta_{\mathbf{G}+\mathbf{G}_i, \mathbf{g}},
\end{equation}
where $\{\mathbf{g}\}\subset\{\mathbf{G}\}$ are
the PC reciprocal lattice vectors.
Applying the operator (\ref{po}) to wave function (\ref{pw}),
and using the Fourier relation (\ref{Fr}) gives
\begin{equation} \label{ppw}
  \mathcal{P}_{\mathbf{K}+\mathbf{G}_i} \psi_\mathbf{K}(\mathbf{x})
  = \sum_\mathbf{g} C_{\mathbf{K}+\mathbf{G}_i-\mathbf{g}}
  e^{i(\mathbf{K}+\mathbf{G}_i-\mathbf{g})\bm{\cdot}\mathbf{x}},
\end{equation}
which is a PC Bloch function with wavevector
$\mathbf{K}+\mathbf{G}_i$.
This proves the first condition.
Applying $\mathcal{P}$ a second time,
and observing that $e^{-i\mathbf{g}\bm{\cdot}{\mathbf{r}}}=1$,
the second condition, $\mathcal{P}^2=\mathcal{P}$, is also proved.
This completes the proof of projective property.

Equation (\ref{ppw}) can also be used in (\ref{sw}) to give the
spectral weights in the plane-wave representation,
\cite{popescu2012}
\begin{equation} \label{wpw}
  W_{\mathbf{K}J}(\mathbf{G}_i)
  = \sum_\mathbf{g} \left| C_{\mathbf{K} + \mathbf{G}_i
      - \mathbf{g}} \right|^2.
\end{equation}
The proof requires only the orthonormality relation,
\begin{equation}
  \int d\mathbf{x} e^{i(\mathbf{g} - \mathbf{g}')\bm{\cdot}\mathbf{x}}
  = \delta_{\mathbf{g}, \mathbf{g}'}.
\end{equation}

Within this representation, it is easy to verify
the following sum rule as well,
\begin{align} \label{sr1}
  \sum_{\mathbf{G}_i} W_{\mathbf{K}J} & =
  \sum_{\mathbf{G}_i} \sum_\mathbf{g}
  \left|C_{\mathbf{K}+\mathbf{G}_i-\mathbf{g}}\right|^2 \nonumber \\
  & = \sum_{\mathbf{G}} \left|C_{\mathbf{K}-\mathbf{G}}\right|^2
    = 1.
\end{align}

Equation~(\ref{wpw}) has a simple interpretation. The set $\{\mathbf{G}\}$
is partitioned into $\mathcal{N}$ sets, and the sums over each
give the unfolded spectral weights.

\begin{figure}
  \includegraphics[width=8.5cm]{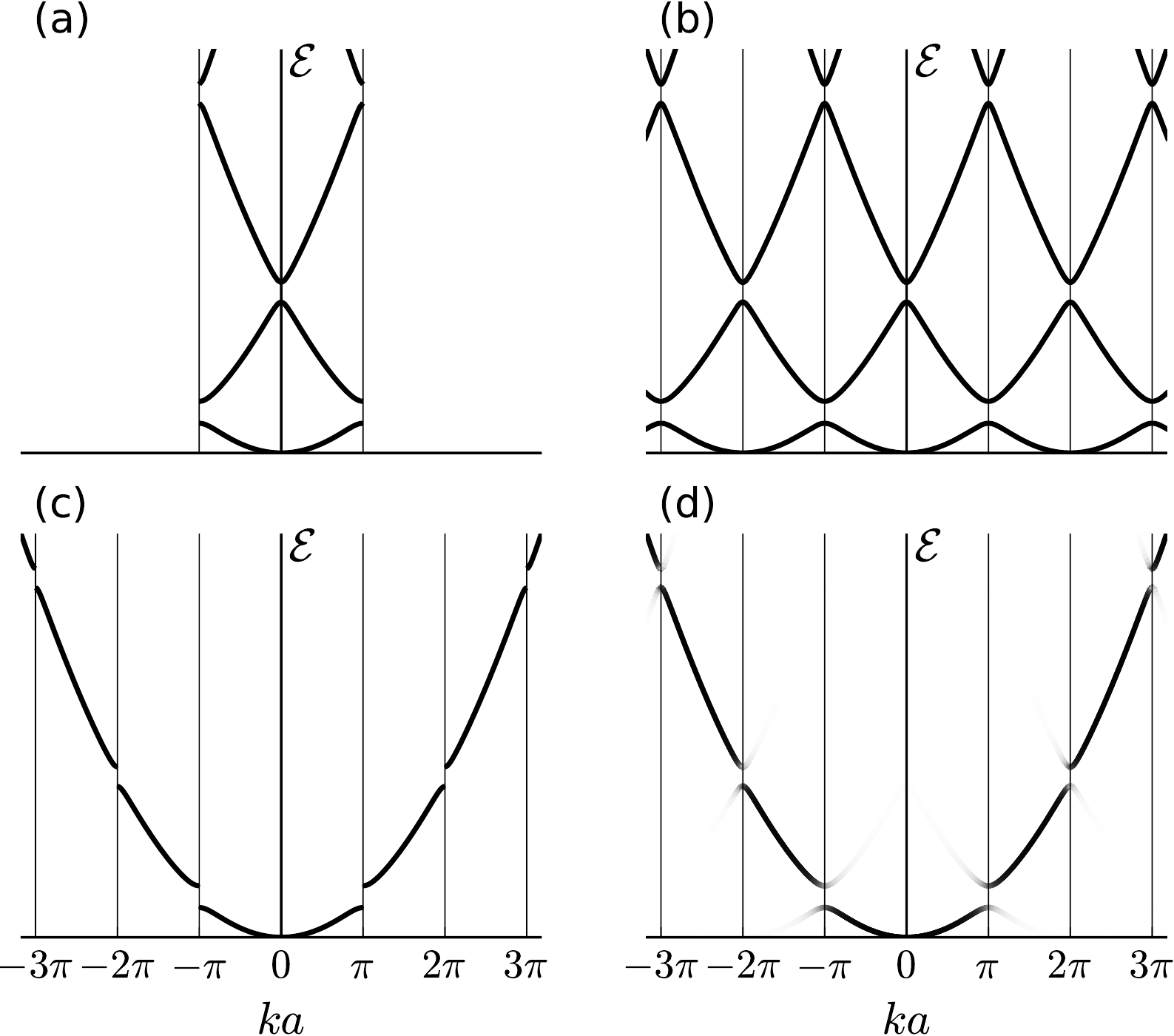}
  \caption{\label{fig1} Band structure of the 1D chain
    in plane wave representation. The size of energy gaps is
    taken to be $5$ units.
    (a) Reduced zone scheme. (b) Repeated zone scheme.
    (c) Extended zone scheme. (d) Unfolded bands.}
\end{figure}

\subsection{Atomic orbital representation}
In this section,
unfolding formulas for localized basis functions are derived.
This covers both the linear combination of atomic orbitals (LCAO),
which uses non-orthogonal basis functions,
and tight-binding models with orthogonal basis functions,
i.e., Wannier functions.
The LCAO method will be treated first
as it covers the tight-binding method as a special case.

The sites of the PC and SC systems are described by $\{\mathbf{r}\}$
and $\{\mathbf{R}\}$, respectively.
In the PC system atomic orbitals are denoted
as $|\mathbf{r}m\rangle$,
with corresponding upper case letters used for the SC system.
Furthermore, an orbital $|\mathbf{R}M\rangle$ in the SC
is described as $|\mathbf{R}+\mathbf{r}(M),m(M)\rangle$ in the PC.
There are $l$ and $L$ unit cells in PC and SC descriptions, respectively,
with $l=\mathcal{N}L$.

An eigenstate of the SC system is given in terms of an
expansion in Bloch functions,
\begin{equation} \label{KJ}
  |\mathbf{K}J\rangle = 
  \sum_N C_N^{\mathbf{K}J} |\mathbf{K}N\rangle,
\end{equation}
where $J$ is a band index and the Bloch functions are given by
\begin{equation} \label{Bf}
  |\mathbf{K}N\rangle
  =\frac{1}{\sqrt{L}} \sum_\mathbf{R}
  e^{i\mathbf{K}\bm{\cdot}\mathbf{R}}
  |\mathbf{R}N\rangle.
\end{equation}
The eigenvectors $C^{\mathbf{K}}$ are solutions
of a generalized eigenvalue problem
\begin{equation}
  H^\mathbf{K}=E^\mathbf{K}S(\mathbf{K})C^\mathbf{K}
\end{equation}
where the overlap matrix $S(\mathbf{K})$ is a Fourier sum of
$\langle0M|\mathbf{R}N\rangle{\equiv}S_{0M,\mathbf{R}N}$ matrix elements
reflecting the non-orthogonality of atomic orbitals.

Projection of an eigenstate (\ref{KJ}) depends on the
projection of a Bloch function (\ref{Bf}),
which in turn depends on the projection of an SC orbital
$|\mathbf{R}N\rangle$,
\begin{align}
  \mathcal{P}_\mathbf{k} |\mathbf{R}M\rangle & =
  \frac{1}{\mathcal{N}} \sum_{j=1}^\mathcal{N}
  e^{-i\mathbf{k}\bm{\cdot}\mathbf{r}_j}
  \mathcal{T}(\mathbf{r}_j)|\mathbf{R}+\mathbf{r}(M),m(M)\rangle \nonumber \\
  & =
  \frac{1}{\mathcal{N}} \sum_{j=1}^\mathcal{N}
  e^{-i\mathbf{k}\bm{\cdot}\mathbf{r}_j}
    |\mathbf{R}+\mathbf{r}(M)-\mathbf{r}_j,m(M)\rangle.
\end{align}
Since $e^{i\mathbf{K}\bm{\cdot}\mathbf{R}}=
e^{i(\mathbf{k}-\mathbf{G}_i)\bm{\cdot}\mathbf{R}}=
e^{i\mathbf{k}\bm{\cdot}\mathbf{R}}$,
\begin{align} \label{PK}
  \mathcal{P}_\mathbf{k} |\mathbf{K}M\rangle & =
  \frac{1}{\mathcal{N}\sqrt{L}} \sum_{\mathbf{R},\mathbf{r}_j}
  e^{i\mathbf{k}\bm{\cdot}(\mathbf{R}-\mathbf{r}_j)}
  |\mathbf{R}-\mathbf{r}_j+\mathbf{r}(M),m(M)\rangle \nonumber \\
  & =
    \frac{1}{\mathcal{N}\sqrt{L}} \sum_{\mathbf{r}}
    e^{i\mathbf{k}\bm{\cdot}\mathbf{r}} |\mathbf{r}+\mathbf{r}(M),m(M)\rangle
    \nonumber \\
  & =
    \frac{1}{\mathcal{N}\sqrt{L}} \sum_{\mathbf{r}}
    e^{i\mathbf{k}\bm{\cdot}[\mathbf{r}-\mathbf{r}(M)]} |\mathbf{r}m(M)\rangle.
\end{align}
Having the projection of
an eigenstate, Eq.~(\ref{PK}), the general spectral weight formula,
Eq.~(\ref{sw}), can be applied to find
\cite{lee2013}
\begin{equation} \label{aosw}
  W_{\mathbf{K}J} = \frac{1}{\mathcal{N}}
  \sum_{NM\mathbf{r}} C_N^{\mathbf{K}J\ast} C_M^{\mathbf{K}J}
  e^{i\mathbf{k}\bm{\cdot}[\mathbf{r}-\mathbf{r}(M)]}
  S_{0N,\mathbf{r}m(M)}.
\end{equation}
In the last step, the periodicity of SC was used to replace
$\sum_{\mathbf{R}}\rightarrow{L}\delta_{\mathbf{R},\mathbf{0}}$.

In orthogonal tight-binding,
the overlap integral becomes
\cite{farjam2014}
\begin{align}
  S_{0N,\mathbf{r}m(M)} & = S_{\mathbf{r}(N)n(N),\mathbf{r}m(M)}
                          \nonumber \\
                        & = \delta_{\mathbf{r}(N),\mathbf{r}}
                          \,\delta_{n(N),m(M)},
\end{align}
so that the spectral weights, Eq.~(\ref{aosw}), simplify to
\begin{multline} \label{tbsw}
  W_{\mathbf{K}J} = \frac{1}{\mathcal{N}}
  \sum_{NM} C_N^{\mathbf{K}J\ast} C_M^{\mathbf{K}J} \\
  \times
  e^{i\mathbf{k}\bm{\cdot}[\mathbf{r}(N)-\mathbf{r}(M)]}
  \delta_{n(N),m(M)}.
\end{multline}

For the LCAO and tight-binding models, in addition to
Eq.~(\ref{sr1}), there is another sum rule given by
\cite{boykin2007}
\begin{equation}
  \sum_J W_{\mathbf{K}J} = \frac{\mathcal{M}}{\mathcal{N}},
\end{equation}
where $\mathcal{M}$ is the number of orbitals in a supercell,
so the right-hand side is just the average number of orbitals
in a primitive cell.
The proof can be understood by noting that a complete basis per PC
must consist of $\mathcal{M}/\mathcal{N}$ orbitals.

\begin{figure}
  \includegraphics[width=8.5cm]{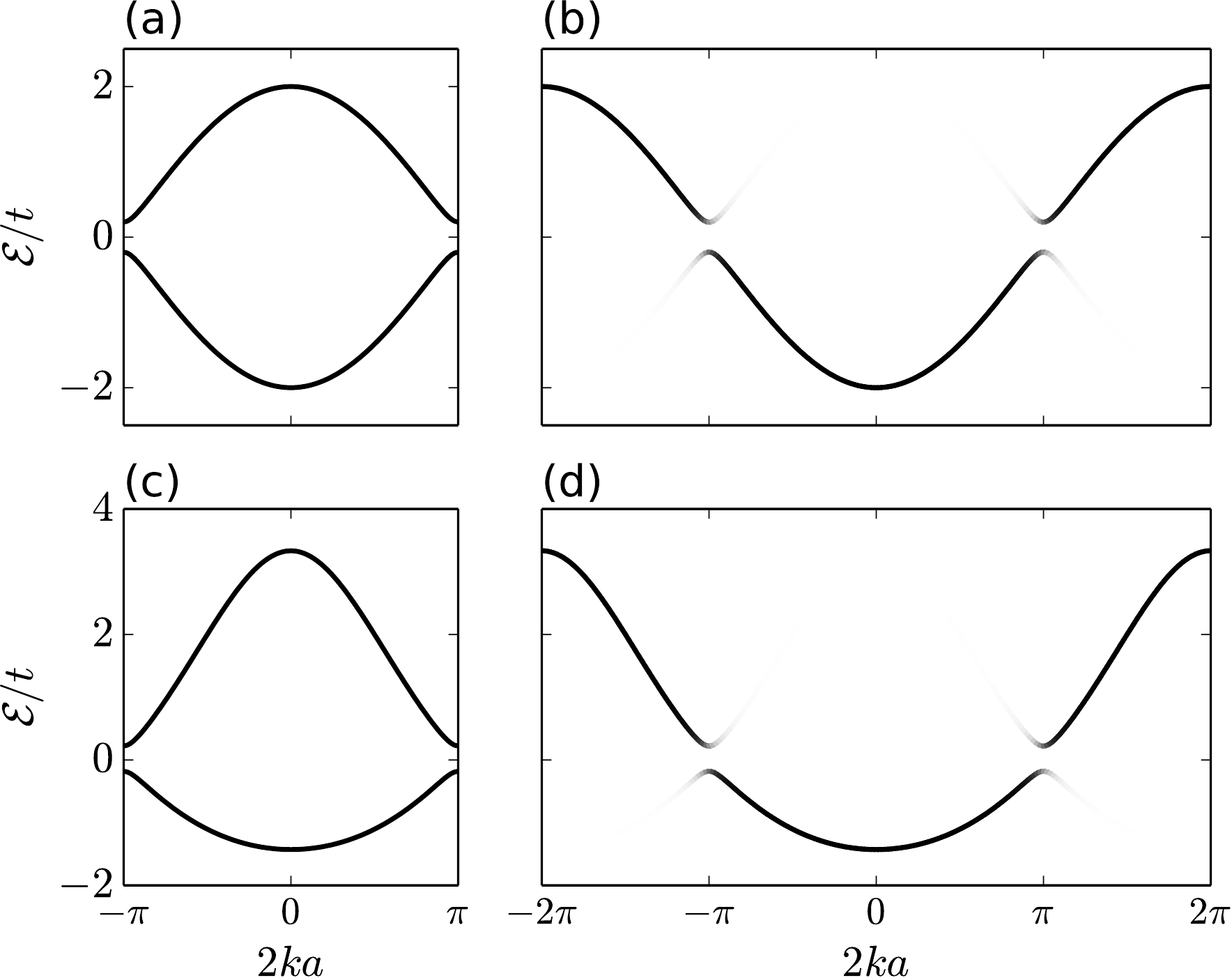}
  \caption{\label{fig2} Band structures of Peierls distorted
    one-dimensional chain. (a) Reduced-zone band structure from
    tight-binding model with the two hoppings
    $t,t'=-0.9,-1.1$. 
    (b) Unfolding bands from same tight-binding model.
    (c) LCAO model with same hoppings as above,
    and overlaps $s,s'=0.15,0.25$.
    (d) LCAO unfolded bands.}
\end{figure}

\section{\label{results}Numerical examples}
A number of toy models are presented in this section to illustrate
the unfolding formulas of Sec.~\ref{theory}.
The examples range from the simplest to illustrate the basic procedure
to the more difficult ones of finite and random systems.

\subsection{Plane-waves in one dimension}
As an illustration of Eq.~(\ref{wpw}),
the simplest one-dimensional (1D) example is considered.

The electrons move in a periodic 1D potential
of period $a$.
The electronic structure is obtained by solving
the Schr\"odinger equation in momentum space
(with $\hbar=2m=1$ units),
\begin{equation} \label{pwse}
  [(K-G)^2 - \mathcal{E}]C_{K-G}
  +\sum_{G'} U_{G'-G} C_{K-G'} = 0,
\end{equation}
where $U_G$ are the Fourier components of the potential,
and $G$ are integer multiples of $b=2\pi/a$.
For small potential the energy spectrum is free-electron-like
with gaps of size $2|U_G|$.
\cite{ashcroft}

The band structure, shown in Fig.~\ref{fig1}, has three equal band gaps,
which is obtained by setting $U_b=U_{2b}=U_{3b}=U$.
The calculation is carried out by truncating
the set of linear equations (\ref{pwse})
to include only $G=0,\pm{b},\pm{2b},\pm{3b},\pm{4b}$,
i.e., a $9\times9$ eigenvalue equation.
Figures~\ref{fig1}(a,b,c) show, respectively, reduced, repeated and extended
zone schemes as found in solid state textbooks.
\cite{ashcroft}
For this simple example the extended zone scheme is obtained
by taking first band in first BZ, the second band in second BZ, and so on.
However, this simple procedure is not feasible for more complicated examples
in higher dimensions.
Figure~1(d) shows the unfolded bands obtained by
considering a hypothetical lattice with a period of $a/4$
and calculating the spectral weights according to (\ref{wpw}).
Here the sum $\sum_g|C_{k-g}|^2$ is applied to the
repeated zone calculation and includes only $g=0,\pm{4b}$.
In contrast to the abrupt changes in the extended zone scheme,
the unfolded bands are observed to diminish smoothly across
the zone boundaries.

\begin{figure}
  \includegraphics[width=8.5cm]{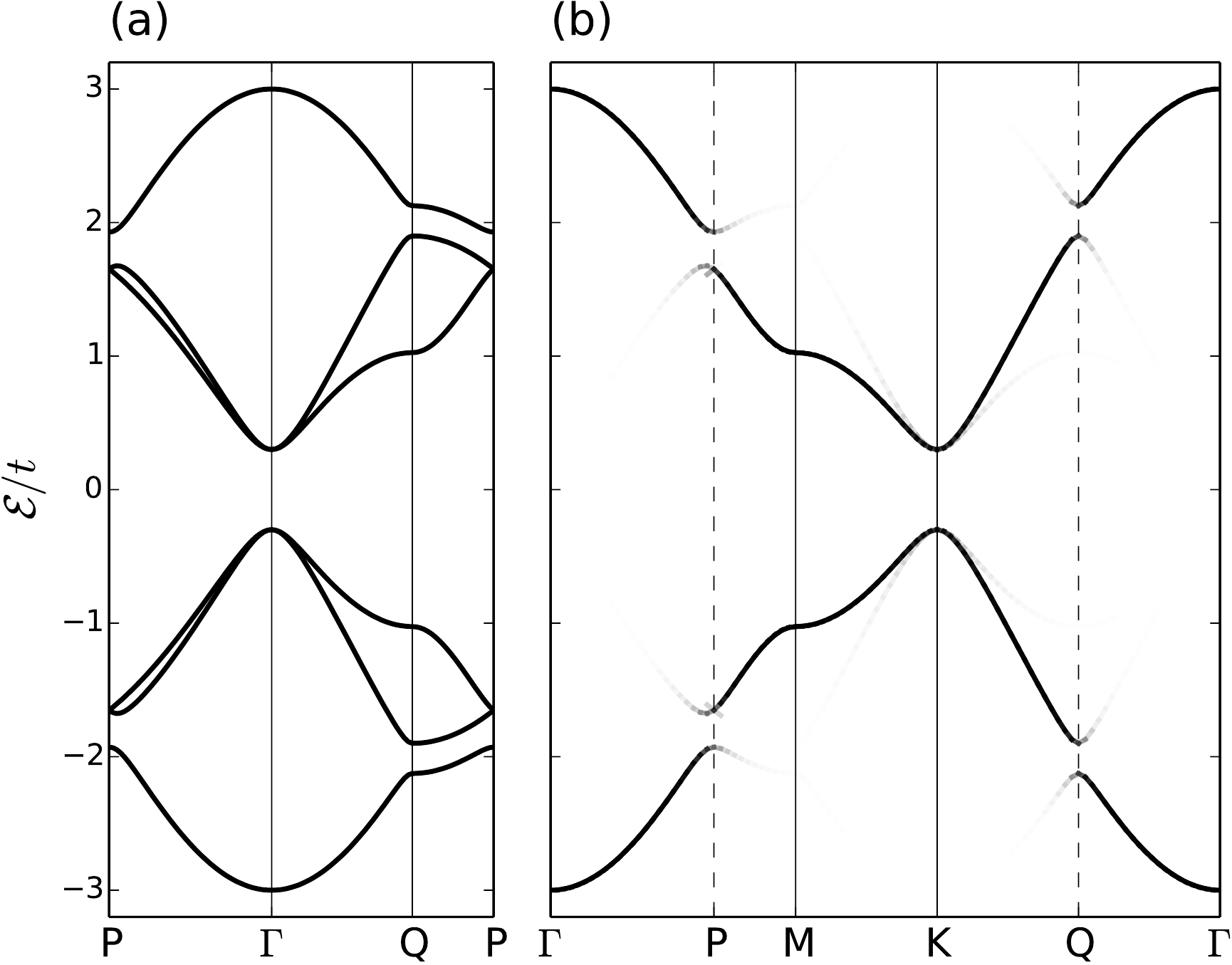}
  \caption{\label{fig3} Band structure of honeycomb lattice with
    Kekul\'e distortion. The hoppings are $t=-1.1$ and $t'=-0.8$ producing
    a band gap of $0.6$ about zero energy. The special points
    $\Gamma$, $K$ and $M$ refer to center, corner and side
    of the hexagonal PBZ, respectively.
    $P$ and $Q$ are special points of SBZ
    and correspond to $K$ and $M$, respectively.
    (a) Reduced zone bands.
    Both $K$ points of the primitive cell Brillouin zone
    are folded onto the $\Gamma$ point of supercell Brillouin zone.
    (b) Unfolded bands showing the gap opening at the $K$ point.
    Smaller gaps have also appeared
    at the zone boundaries of supercell Brillouin zone.}
\end{figure}

\subsection{1D tight-binding model}
In this example, the 1D chain is assumed to have
undergone a Peierls distortion,
which doubles the period to $2a$.
The nearest-neighbor tight-binding model consists of
two alternate hoppings $t$ and $t'$,
and the transfer integral matrix is given by
\begin{equation}
  H^K = \left(
    \begin{matrix}
      0 & t' + t\,e^{-2iKa} \\
      t' + t\,e^{2iKa} & 0
    \end{matrix} \right).
\end{equation}
The two energy bands are shown in Fig.~\ref{fig2}(a).

These bands can be unfolded to the BZ of the underlying cell
of period $a$ via the spectral weights (\ref{tbsw}).
Writing the eigenvector coefficients
$C^{KJ}_N$ simply as $C_N$ the spectral weight of a band
is given for this example by
\begin{equation} \label{xtbw}
  W = \frac{1}{2}(C_1^\ast+C_2^\ast e^{ika})
  (C_1+C_2 e^{-ika}).
\end{equation}
The corresponding unfolded bands are shown in Fig.~\ref{fig2}(b).

\begin{figure}
  \includegraphics[width=8.5cm]{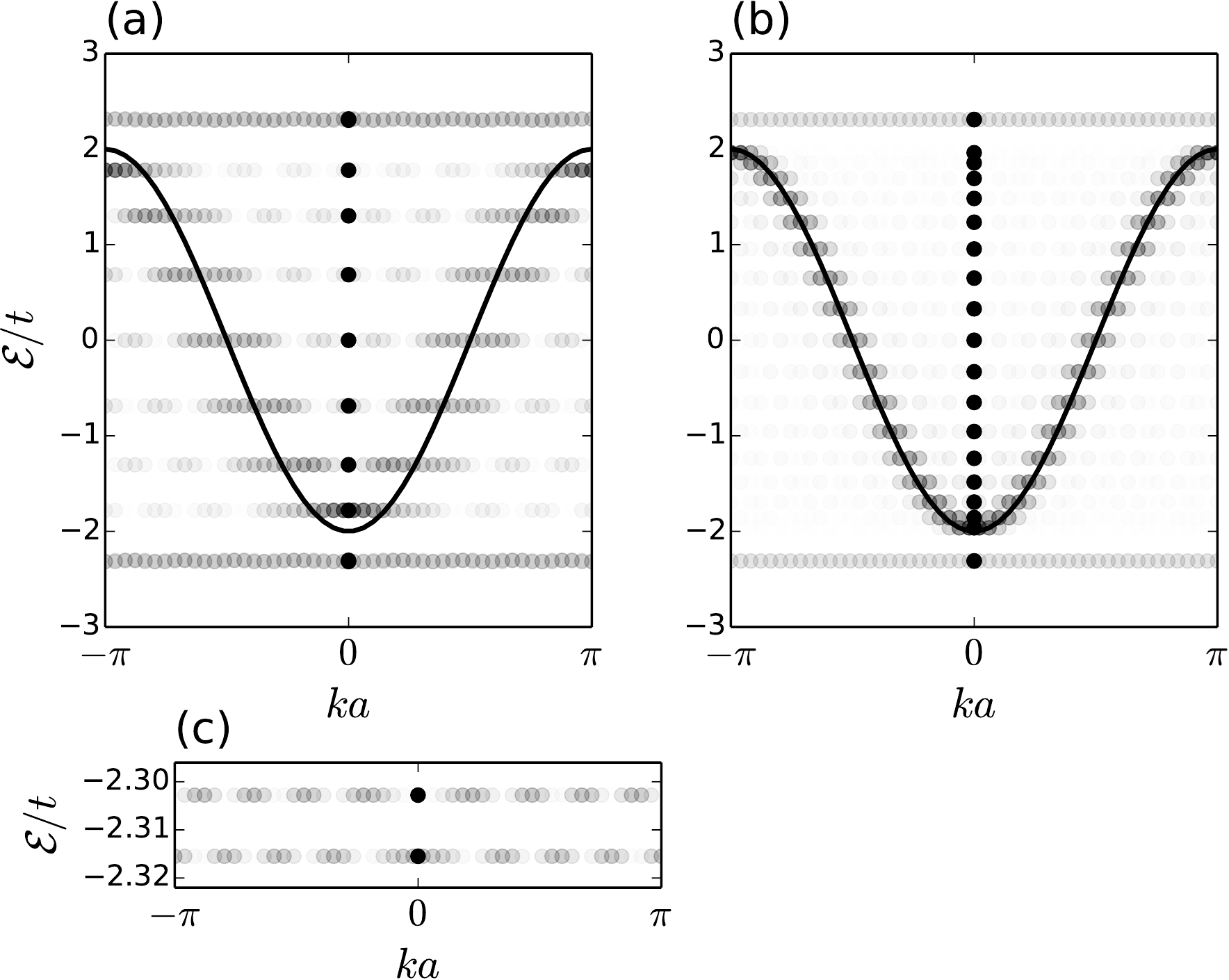}
  \caption{\label{fig4} Energy spectrum of tight-binding electrons
    on a finite chain. (a) Chain of $11$ atoms. Hoppings at both ends
    are enhanced by a factor of $2$.
    Black points at $K=0$ represent the eigenvalues,
    and the black curve is the exact dispersion
    of the infinite chain, $E=2t\cos(ka)$.
    The $K=0$ point has been unfolded to $50$ equally spaced $k$ points
    throughout the zone.
    The levels below and above the band limits
    each represent the overlap of two
    nearly degenerate eigenvalues,
    which can be resolved on a finer energy scale (c).
    These are localized edge states and, thus,
    their unfolded points are extended throughout the zone.
    (b) Chain of $21$ atoms. Increasing the number of atoms has made
    the unfolded spectrum to become more similar
    to the infinite chain band.
    Similary to the $11$ atom chain, the edge levels are doubly degenerate
    on this scale.}
\end{figure}

\subsection{1D LCAO model}
The LCAO model is similar to the tight-binding model
but in addition to the hoppings $t$ and $t'$ there are
overlap integrals $s$ and $s'$.
This creates an asymmetry in the band structure.
The overlap integral matrix to be used
in the generalized eigenvalue equation is
\begin{equation}
  S(K) = \left(
    \begin{matrix}
      1 & s' + s\,e^{-2iKa} \\
      s' + s\,e^{2iKa} & 1
    \end{matrix} \right).
\end{equation}
The resulting bands are shown in Fig.~\ref{fig2}(c).

The unfolded spectral weights, Eq.~(\ref{aosw}),
can be written in a similar form as Eq.~(\ref{xtbw}) as
\begin{equation}
  W = \frac{1}{2}(C_1^\ast A+C_2^\ast e^{ika} A^\ast)
  (C_1+C_2 e^{-ika}),
\end{equation}
where $A=1+s\,e^{-ika}+s'\,e^{ika}$.
The unfolded bands are shown in Fig.~\ref{fig2}(d).

\subsection{Honeycomb lattice with Kekul\'e distortion}
In the two-dimensional
honeycomb lattice bond alternation is
known as Kekul\'e distortion,
and has an important implication for the energy spectrum.
\cite{hou2007}
The structure can be described by a commensurate lattice
related to the underlying lattice by a nonsingular transformation matrix
with integer elements,
\begin{equation}
  \left( \begin{matrix}
      \mathbf{A}_1 \\
      \mathbf{A}_2
    \end{matrix} \right) =
  \left( \begin{matrix}
      2 & 1 \\
      1 & 2
    \end{matrix} \right)
  \left( \begin{matrix}
      \mathbf{a}_1 \\
      \mathbf{a}_2
      \end{matrix} \right).
\end{equation}
The new unit cell is a $(\sqrt{3}\times\sqrt{3})R30^\circ$
supercell with $3$ times the area of the primitive unit cell.
(See Ref.~[\onlinecite{farjam2009}] for more details.)
There are two sites, or basis orbitals, per primitive unit cell
and six sites per supercell.

The Kekul\'e distortion induces
a gap in the spectrum of honeycomb lattice,
given by $E_g=2|t-t'|$,
where $t$ and $t'$ are the two hoppings.
The transfer integral matrix is given by
\begin{widetext}
  \begin{equation}
    H^\mathbf{K} =
    \left( \begin{matrix}
        0 & t & 0 & t & 0 & t'e^{-i\mathbf{K}\bm{\cdot}\mathbf{A}_3} \\
        t & 0 & t & 0 & t'e^{-i\mathbf{K}\bm{\cdot}\mathbf{A}_1} & 0 \\
        0 & t & 0 & t'e^{i\mathbf{K}\bm{\cdot}\mathbf{A}_2} & 0 & t \\
        t & 0 & t'e^{-i\mathbf{K}\bm{\cdot}\mathbf{A}_2} & 0 & t & 0 \\
        0 & t'e^{i\mathbf{K}\bm{\cdot}\mathbf{A}_1} & 0 & t & 0 & t \\
        t'e^{i\mathbf{K}\bm{\cdot}\mathbf{A}_3} & 0 & t & 0 & t & 0
      \end{matrix} \right).
  \end{equation}
\end{widetext}

The band structure, in reduced zone and unfolded extended zone schemes,
respectively, are shown in Fig.~\ref{fig3}.
This result is particularly graphic in showing the usefulness of unfolding.
While the SC folded bands have little resemblance to the bands
of the perfect system,
the unfolded effective band structure is quite comparable to
the unperturbed one, different only in band gaps at the $K$ point
and the zone boundaries.

\subsection{Finite 1D chain of atoms}
An important application of unfolding concerns
the electronic structure of surface layers.
\cite{allen2013, chen2014}
Commonly surface states of a crystal are studied by using a slab
of several layers of atoms and forming a supercell by the addition of
a vacuum layer,
thick enough to prevent interaction between the slabs.
For the vertical component, only the $K=0$ point needs to be considered.

The slab approach is illustrated here with a finite 1D chain of atoms.
The Hamiltonian matrix of a nearest-neighbor tight-binding model of electrons
in a finite one-dimensional chain is given by
\begin{equation}
  \bm{\mathsf{H}} = \left( \begin{matrix}
      0 & t' & 0 & \dots & 0 & 0 & 0 \\
      t' & 0 & t & \dots & 0 & 0 & 0 \\
      0 & t & 0 & \dots & 0 & 0 & 0 \\
      \vdots & \vdots & \vdots & \ddots & \vdots & \vdots & \vdots \\
      0 & 0 & 0 & \dots & 0 & t & 0 \\
      0 & 0 & 0 & \dots & t & 0 & t' \\
      0 & 0 & 0 & \dots & 0 & t' & 0
    \end{matrix} \right),
\end{equation}
where the hoppings for the end atoms are allowed to be different.
The eigenvalue for this Hamiltonian is solved to give the energy spectrum
and eigenvectors,
which correspond to $K=0$ in a supercell calculation.
The results are shown for chains of $11$ and $21$ atoms in Fig.~\ref{fig4}.
The $K=0$ point unfolds into a number $\mathcal{N}$ (size of SC) of $k$ points.
The weights are seen to be smeared in $k$ as a result
of the chain being finite, but most of the weights are concentrated
near the dispersion curve of the infinite system.
As the chain increases its length and approximates the infinite one better,
spectral broadening in $k$ is seen to decrease.
There are two nearly degenerate states below, and two above, the band
limits, which are smeared throughout the BZ.
These are localized end point states.

\begin{figure}
  \includegraphics[width=8.5cm]{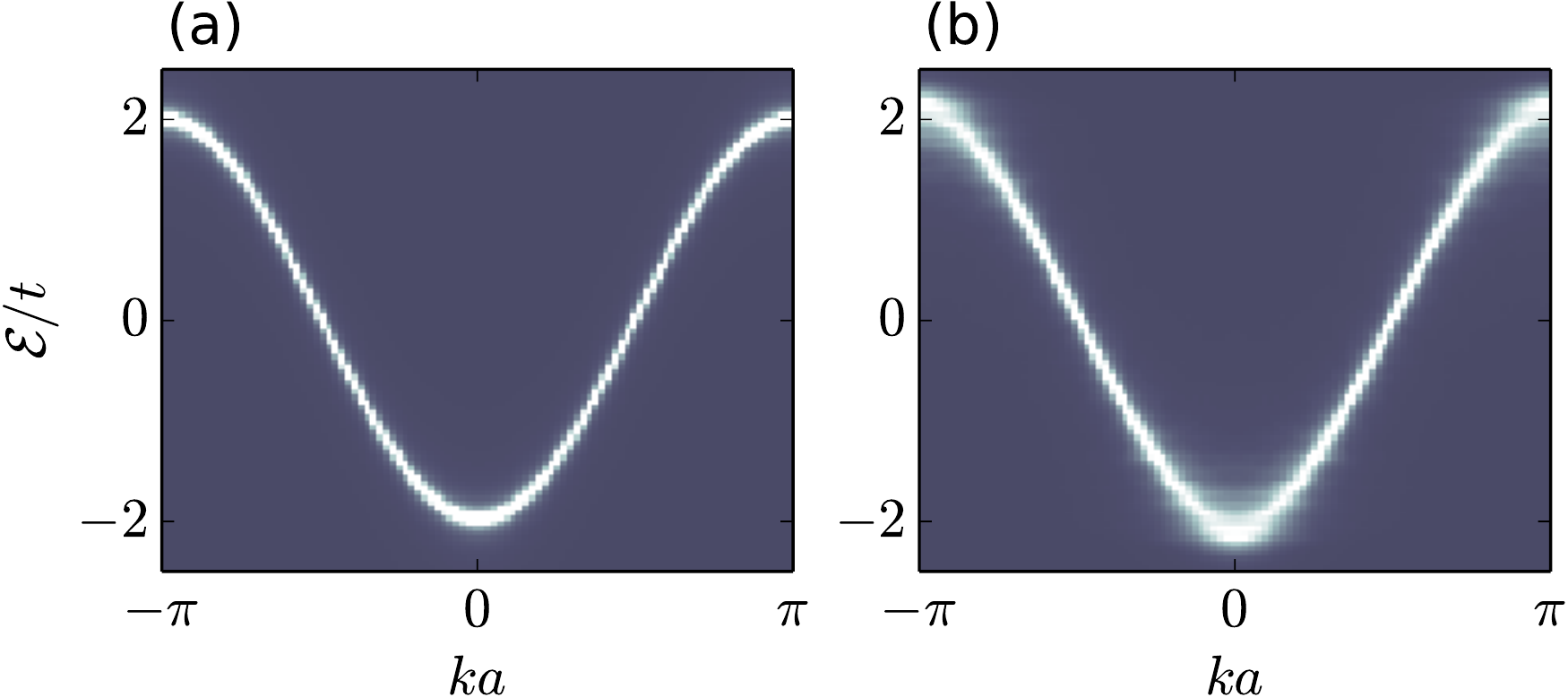}
  \caption{\label{fig5} Unfolded band structures of a 1D chain described
    by a supercell of $128$ atoms.
    An `intrinsic' broadening of $\eta=0.03$ has been used
    in order to obtain smooth spectral functions.
    (a) Perfect system with hopping $t=-1$.
    (b) Disordered system with equal numbers of two hoppings given by
    $t=-1.2$ and $t'=-0.8$. The spectral function was obtained by
    averaging over $12$ random samples.}
\end{figure}

\subsection{Disordered 1D chain}
Another important application of unfolding is to obtain effective band
structures of random alloys,
which requires very large supercells and is computationally expensive.
\cite{boykin2007, popescu2012}
This is illustrated here by the 1D chain of atoms
with two different hoppings distributed randomly.
The approach used is to average the unfolded spectral functions of
a set of random realizations of the system with a relatively small supercell.
Alternatively, one can use one random realization of a very large supercell.
\cite{popescu2012}
The unfolded spectral function for the 1D model is given by
\begin{equation}
  A(k, E) = \sum_J W_{KJ} \delta(E - E_{KJ}),
\end{equation}
where $W_{KJ}$ are the unfolded spectral weights given by Eq.~(\ref{tbsw}).
In the numerical calculations,
the delta function is represented by a Lorentzian with
width parameter $\eta$.
The calculated spectral functions of perfect and disordered chain
are shown in Fig.~\ref{fig5},
where the main effect of disorder is seen as a broadening
in both $k$ and $E$.

\section{Conclusions}
Although band structure unfolding formulas can already be
used as a computational tool,
it is desirable to have a clear and intuitive understanding
of the concepts behind them.
This revisit makes a contribution toward this aim by presenting a
simple and unified description of the unfolding method
based on the $k$-projection operator.
In addition, a range of simple examples are presented in pedagogical style
that demonstrate interesting aspects of the unfolding method.

\begin{acknowledgments}
  The author thanks Afshin Namiranian for useful discussions,
  and IPM for finanical support.
\end{acknowledgments}

\bibliography{main}

\end{document}